\documentclass[aps,prc,reprint,nofootinbib]{revtex4-1}
\usepackage{graphicx}   
\usepackage{latexsym}   
\usepackage{enumerate}
\usepackage{tabularx}

\begin{document}
\title{Directed, elliptic and triangular flow of protons in Au+Au reactions at 1.23 A~GeV:\\ A theoretical analysis of the recent HADES data}

\author{Paula Hillmann$^{1,2}$, Jan Steinheimer$^2$,  Marcus~Bleicher$^{1,2,3,4}$}

\affiliation{$^1$ Institut f\"ur Theoretische Physik, Goethe Universit\"at Frankfurt, Max-von-Laue-Str. 1, D-60438 Frankfurt am Main, Germany}
\affiliation{$^2$ Frankfurt Institute for Advanced Studies, Ruth-Moufang-Str. 1, 60438 Frankfurt am Main, Germany}
\affiliation{$^3$ GSI Helmholtzzentrum f\"ur Schwerionenforschung GmbH, Planckstr. 1, 64291 Darmstadt, Germany}
\affiliation{$^4$ John von Neumann-Institut f\"ur Computing, Forschungzentrum J\"ulich,
52425 J\"ulich, Germany}


\begin{abstract}
Recently, the HADES experiment at GSI has provided preliminary data on the directed flow, $v_1$ elliptic flow, $v_2$ and triangular flow, $v_3$ of protons in Au+Au reactions at a beam energy of 1.23 A GeV. Here we present a theoretical discussion of these flow harmonics within the UrQMD transport approach. We show that all flow harmonics, including the triangular flow, provide a consistent picture of the expansion of the system, if potential interactions are taken into account. Investigating the dependence of the flow harmonics on the nuclear interaction potentials it is shown that especially $v_3$ can serve as a sensitive probe for the nuclear equation of state at such low energies. The triangular flow and its excitation function with respect to the reaction-plane were calculated for the first time and indicate a complex interplay of the time-evolution of the system and the initial conditions at low beam-energies. Our study also indicates a significant softening of the equation of state at beam energies above $E_{\mathrm{lab}}> 7$A GeV which can be explored by at the future FAIR facility.
\end{abstract}

\maketitle

\section{Introduction}

The collision of heavy and light ions at various beam energies allows to explore the properties and dynamics of strongly interacting matter, i.e. matter governed by the laws of Quantum Chromodynamics (QCD), at varying densities and temperatures. QCD-matter under extreme conditions has been present during the first micro seconds after the Big Bang and can be found in neutron stars and other compact stellar objects. To obtain insights into the properties of QCD-matter at 3-4 times the nuclear ground state density and at moderate temperatures the HADES experiment has recently performed collisions of gold nuclei at a beam energy of 1.23 A GeV \cite{Kardan:2016uog}. This density and temperature range is particularly interesting as one expects the transition from dense nuclear matter to dense deconfined quark matter to occur just above this density range. Even more exotic forms of matter, like quarkyonic \cite{McLerran:2008ux} or color superconducting matter \cite{Ruester:2006aj} have been predicted to exist in the density regime under investigation. \par
An established method to study the equation of state of QCD-matter in nuclear collisions is to investigate the development of collective flow. The collective motion of observable hadrons is expected to be a sensitive probe to pressure gradients and inter-particle potentials during the dense phase of the reaction. In particular the so called $v_n$'s, the expansion coefficients of the transverse momentum distribution as Fourier series \cite{Poskanzer:1998yz}:
\begin{equation}
E \frac{{\mathrm d}^3 N}{{\mathrm d}^3 {p}} = 
\frac{1}{2\pi} \frac {{\mathrm d}^2N}{p_{\mathrm{T}}{\mathrm d}p_{\mathrm T}{\mathrm d} y} 
  \left(1 + 2\sum_{n=1}^{\infty}  v_n \cos[n(\varphi-\Psi_{\rm RP})]\right), 
\label{fs}
\end{equation}
are of interest regarding the study of the bulk matter equation of state \cite{Stoecker:1979mj,Hofmann:1976dy,Stoecker:1986ci,Nara:2017qcg,Nara:2016phs,Isse:2005nk,Petersen:2006vm,Voloshin:2002wa,Snellings:2011sz,Ollitrault:1997vz,Yan:2013laa,Retinskaya:2012ky,Yin:2017qhg,Ivanov:2014ioa,Nara:2016hbg}.
Here $\Psi_{\rm RP}$ denotes the reaction plane angle and $\varphi$ is the azimuthal angle with respect to the reaction plane. Note that in our simulations we always set $\Psi_{\rm RP}=0$, as given by the initial geometry of the reaction. 
The coefficients $v_n$ can then be readily calculated using \cite{Poskanzer:1998yz}:
\begin{equation}
v_n(p_{\mathrm{T}},y) = \langle \cos[n\varphi] \rangle,
\label{fc}
\end{equation}
where the average runs over all particles in a given event and acceptance as well as over all events in a given centrality class.

Until now higher order ($n>2$) Fourier coefficients have only been investigated at high beam energies, i.e. at the RHIC and LHC, and are usually connected to initial state fluctuations which are not correlated with the reaction plane of the event \cite{Adam:2016nfo,Esumi:2017qof,He:2017qsk,Krzewicki:2011ee,Alver:2010gr}. Recently the HADES experiment at the SIS18 accelerator has begun studying moments of the azimuthal momentum distribution at a rather low beam energy and with respect to the measured reaction plane. Since the HADES experiment has collected a large amount of data, higher order $v_n$'s can be extracted which may open new possibilities on the study of the properties of dense and moderately hot QCD matter.

In this paper we present results on the flow coefficients $v_n$ of protons, at the HADES beam energy. The dependencies of the directed flow, $v_1$ elliptic flow, $v_2$ and triangular flow, $v_3$, both on rapidity and transverse momentum are studied. To this aim, we employ the UrQMD transport model \cite{Bass:1998ca,Bleicher:1999xi}. 

\section{The UrQMD model}
The Ultra relativistic Quantum Molecular Dynamics (UrQMD) transport model is based on the binary elastic and inelastic scattering of hadrons, including resonance excitations and decays, as well as string dynamics and strangeness exchange reactions \cite{Bass:1998ca,Bleicher:1999xi,Graef:2014mra}. The model employs a geometrical interpretation of scattering cross sections which are taken, when available, from experimental data \cite{Patrignani:2016xqp} or model calculations, e.g. the additive quark model or meson exchange models. In our investigations we use the most recent version of the UrQMD model in its cascade version and compare it to the simulations including soft and hard hadronic potentials. In all versions of UrQMD particle production occurs through the intermediate excitation and decay of resonances or color strings. At high beam energies $\sqrt{s_{\mathrm{NN}}}>8$ GeV, the cascade version of the model successfully explains the mean particle production as well as collective flow in nuclear collisions \cite{Bleicher:2005ti,Petersen:2006vm}. At lower beam energies the role of long range nuclear and electromagnetic interactions becomes more important. Therefore potential interactions have been included in previous versions of the UrQMD code \cite{Bass:1998ca,Hartnack:1997ez}.

\subsection{Potentials in UrQMD}
\begin{figure}[t]	
\includegraphics[width=0.5\textwidth]{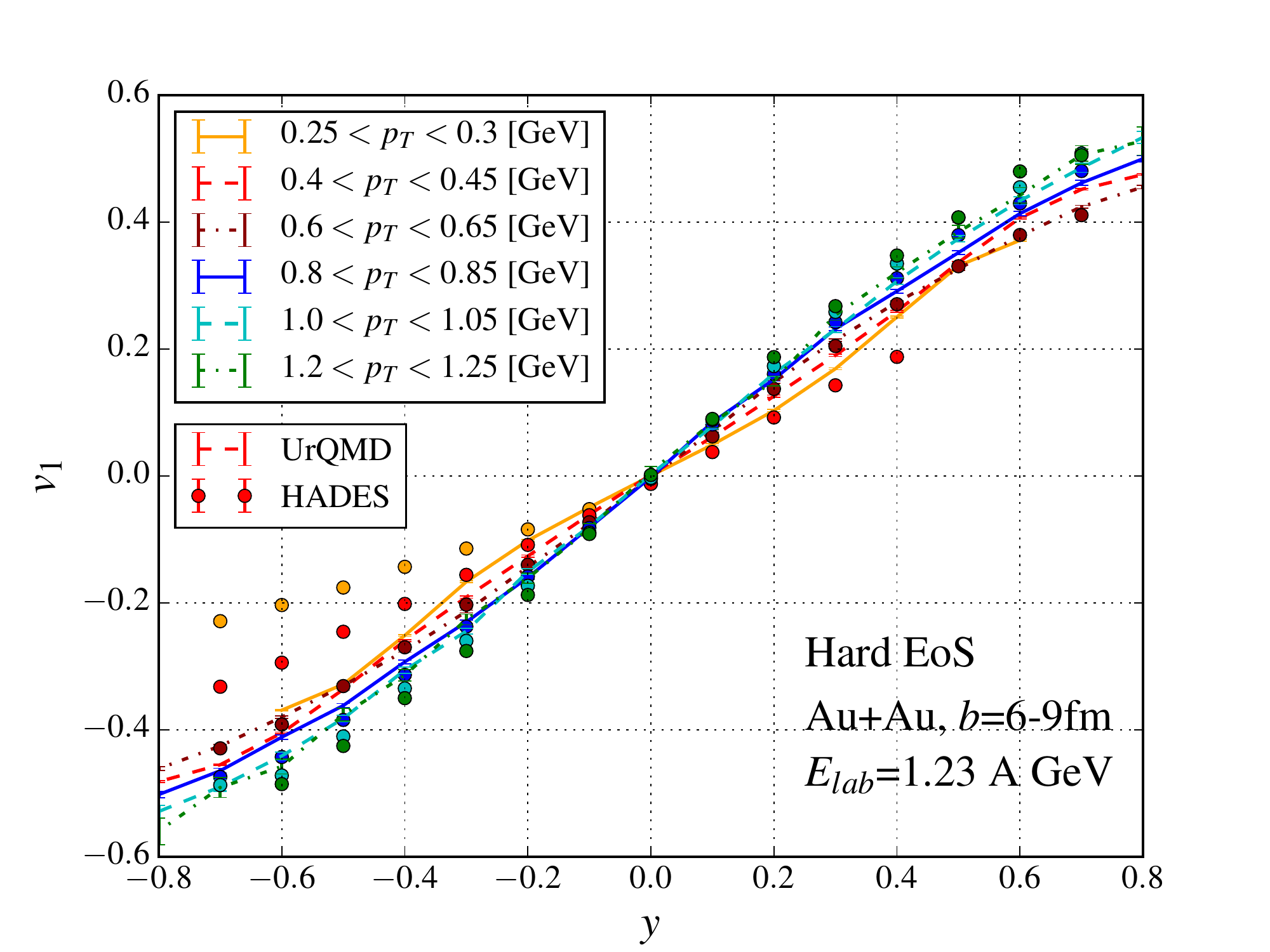}
\caption{[Color online] Directed flow of protons in Au+Au reactions (calculation: $b=6-9$ fm, data: 20\%-30\% centrality) as a function of rapidity and for various transverse momentum regions. The symbols denote the preliminary experimental data of the HADES collaboration~\cite{Kardan:2017knj}, the lines indicate the UrQMD calculations with the hard equation of state. The rapidity is given in the cm-frame.}\label{f1}
\end{figure}		

\begin{figure}[t]	
\includegraphics[width=0.5\textwidth]{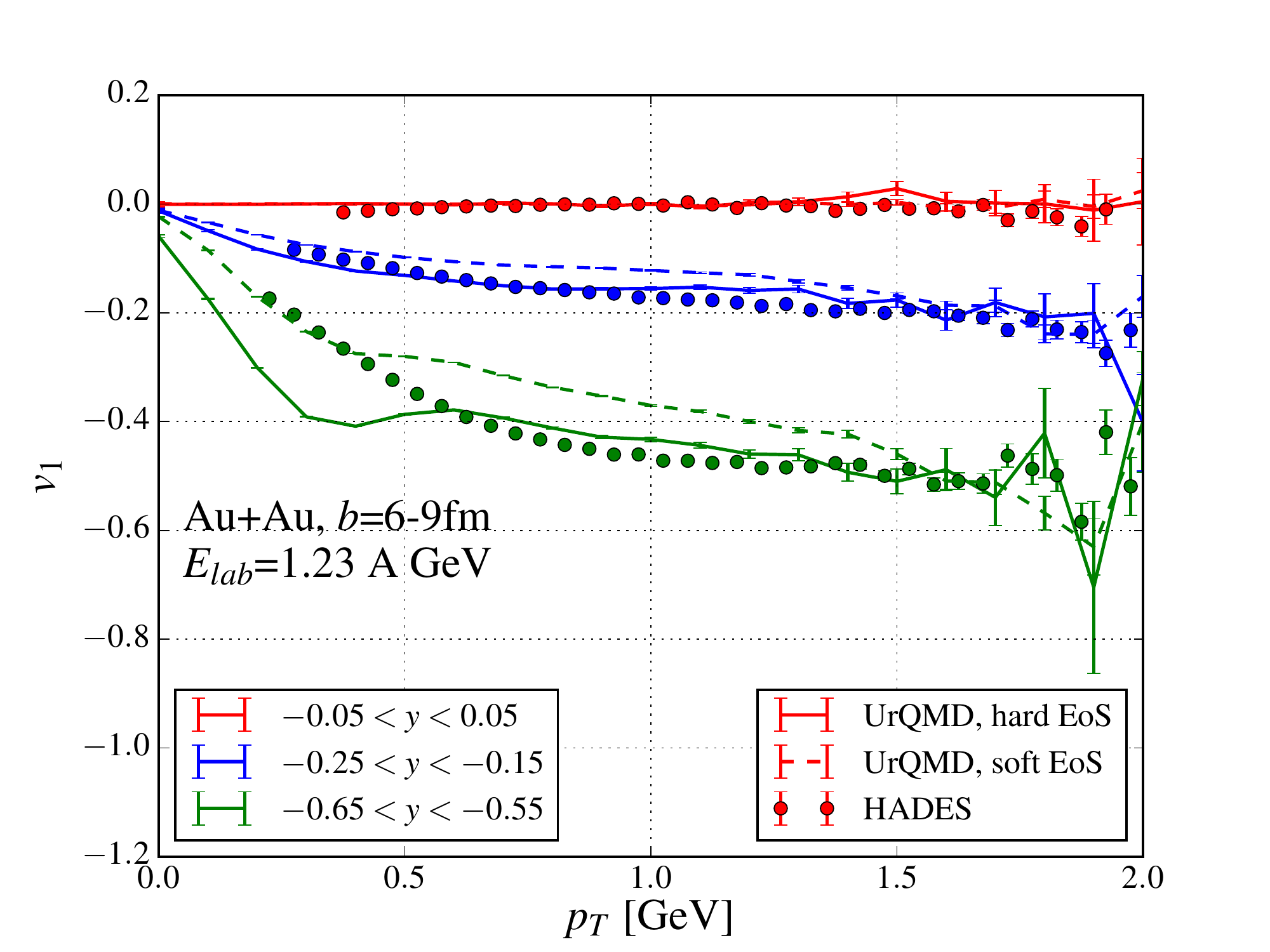}
\caption{[Color online] Directed flow of protons in Au+Au reactions (calculation: $b=6-9$ fm, data: 20\%-30\% centrality) as a function of transverse momentum and for various rapidity regions. The symbols denote the preliminary experimental data by the HADES collaboration~\cite{Kardan:2017knj}, the lines indicate the UrQMD calculations(full lines: hard EoS, dashed lines: soft EoS) for three different rapidity windows. }\label{f2}
\end{figure}		

In the current public version of the UrQMD model (v3.4) the same potential models as in the QMD model are implemented. The long range interactions between electric charges is described by a Coulomb potential.
The Coulomb-potential, $V^{ij}_C$, is given in form of two-particle interactions, where $Z$ is the charge number of the corresponding particles, e is the elementary charge~\cite{Bass:1998ca}:
\begin{equation}
\centering
V^{ij}_C=\frac{Z_iZ_je^2}{\left| \bf{r}_i-\bf{r}_j \right|}
\end{equation}
and $\left| \bf{r}_i-\bf{r}_j \right|$ is the distance between two particles in the center-of-mass frame.

The Yukawa-potential, $V^{ij}_Y$, is given in form of two-particle interactions, where $V^Y_0=-0.498$ MeV fm and $\gamma _Y=1.4$ fm ~\cite{Bass:1998ca}:
\begin{equation}
\centering
V^{ij}_Y=V^Y_0\cdot \frac{\exp{\left( \left| \bf{r}_i-\bf{r}_j \right|/ \gamma _Y\right)}}{\left| \bf{r}_i-\bf{r}_j \right|}
\end{equation}

The hadronic Skyrme-potential, $V_{Sk}$, which defines the stiffness of the EoS is given by~\cite{Hartnack:1997ez}:\\
\begin{equation}
\centering
V_{Sk}=\alpha \cdot \left( \frac{\rho _{int}}{\rho _0}\right) +\beta \cdot \left( \frac{\rho _{int}}{\rho _0}\right) ^{\gamma}
\end{equation}
with $\rho _{int}$ being the baryon density and $\rho _0$ being the ground state  baryon density.\\

\begin{table}[h!]
\centering
\begin{tabular}{|c|c|c|}
\hline
Parameters & hard EoS & soft EoS \\
\hline
$\alpha$ [MeV] & -124 & -356 \\
\hline
$\beta$  [MeV] & 71 & 303 \\
\hline
$\gamma$ & 2.00 & 1.17\\
\hline
\end{tabular}
\caption{Parameters used in the UrQMD  Skyrme potential~\cite{Hartnack:1997ez}.\label{t1}}
\end{table}

By changing the parameters $\alpha$, $\beta$ and $\gamma$ one changes the stiffness, i.e. effective speed of sound, of the nuclear equation of state. In the following we will use two parameterizations, denoted 'hard' and 'soft' equation of state.
The parameters which are implemented for both equations of state are shown in table \ref{t1}, and have been discussed in \cite{Hartnack:1997ez}.

Note that at the moment we do not include effects of momentum or iso-spin dependent potentials which have been discussed in the literature \cite{Li:2005gfa,Li:2007yd,Li:2005zza}. 

\begin{figure}[t]	
\includegraphics[width=0.5\textwidth]{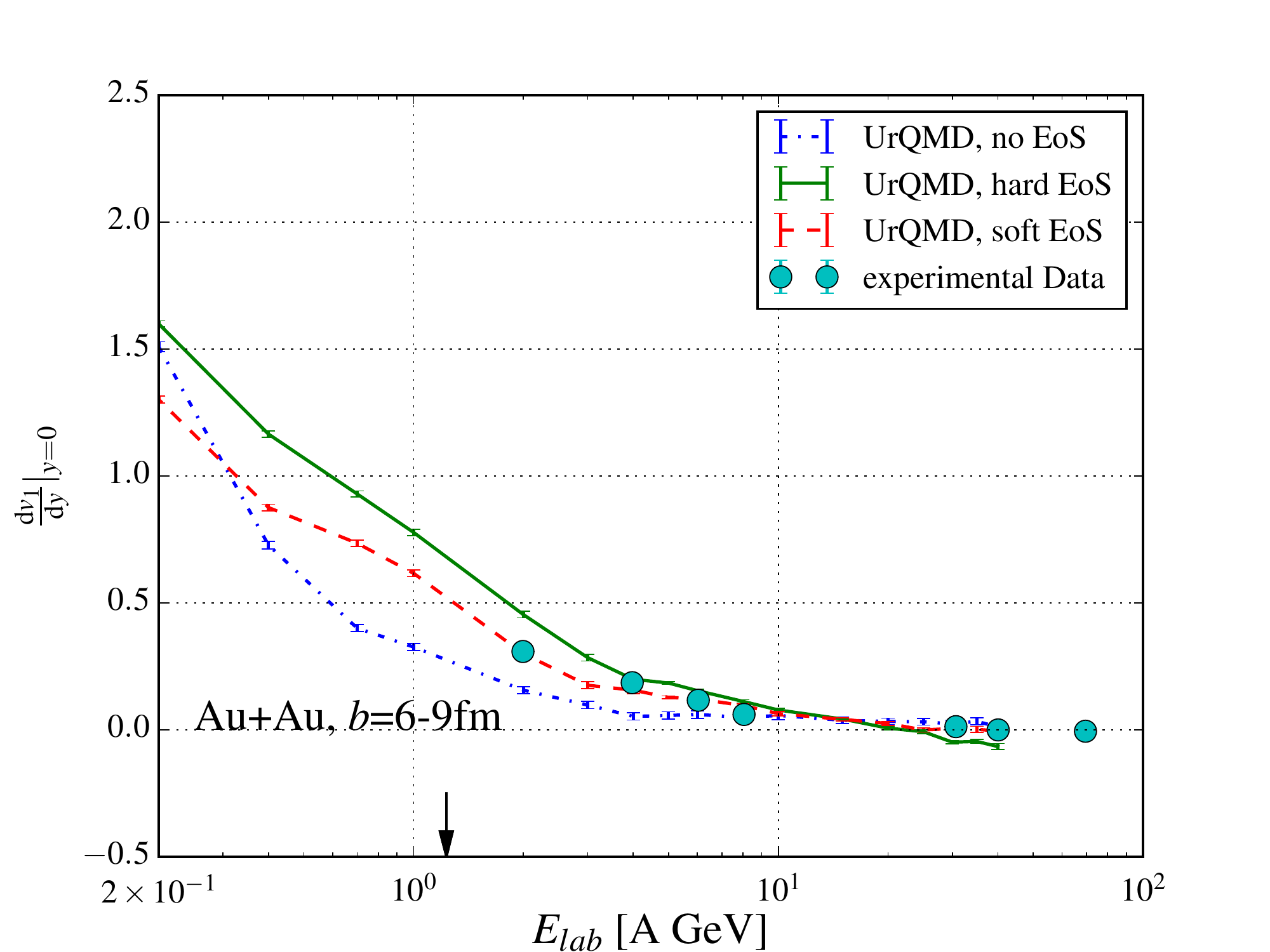}
\caption{[Color online] Excitation function of the midrapidity slope parameters of $v_1$ of protons in Au+Au collisions (calculation: $b=6-9$ fm, data: 20\%-30\% centrality) for the cascade calculation and a soft and hard equation of state. The symbols denote the experimental data \cite{Alt:2003ab, Liu:2000am, Adamczyk:2014ipa}, the lines indicate the UrQMD calculations. The arrow remarks the HADES energy.}\label{f3}
\end{figure}		

\section{results}

In the following we present results for mid-peripheral Au+Au collisions, simulated with the UrQMD transport model. The calculated centralities are $b=6-9$ fm, corresponding to the 20\%-30\% centrality class in the data. In line with the recent HADES data we focus on the flow of protons. All the following results are shown for protons at kinetic freeze-out i.e. at the point in time of their last scattering. To reduce the influence of spectator protons we use a spectator-cut which removes particles with $p_T \leq 0.3$ GeV and $y\geq 0.6$. For the present study we ignore the formation of nuclear clusters from protons and neutrons.

For the rapidity dependence of $v_1$ and $v_2$ at HADES energy only the hard equation of state is considered, while for the $p_T$ dependence and excitation functions of the different flow components and the rapidity dependence of $v_3$ both EoS will be compared to experimental data and UrQMD in its cascade mode.

\subsection{Directed flow, $v_1$}

Figure \ref{f1} shows the directed flow in Au+Au collisions (calculation: $b=6-9$ fm, data: 20\%-30\% centrality) at a fixed target beam energy of $E_{\mathrm{lab}}= 1.23$ A GeV as a function of rapidity and for various transverse momentum intervals. The symbols denote the preliminary experimental data~\cite{Kardan:2017knj}, the lines indicate the UrQMD calculations with a hard EoS. As expected for such a low beam energy the slope of the proton $v_1$ is positive. Also no strong dependence on the transverse momentum is observed, at least above momenta of $p_T>0.5$ GeV. For low momenta and large rapidities we observe a deviation from the measured $v_1$ values. This deviation might be due to the lack of cluster/nuclei formation in our current model implementation. The low transverse momentum region at rapidity close to the beam rapidity may be dominated by nuclear clusters formed from projectile/target fragments. For small rapidity $|y|<0.2$ the model describes the data very well.

Figure \ref{f2} shows the directed flow in Au+Au collisions (20\%-30\% centrality) as a function of transverse momentum for various rapidity regions. The symbols denote the preliminary experimental data~\cite{Kardan:2017knj}, the lines indicate the UrQMD calculations. Again we observe a very good description of the data for the hard EoS and transverse momenta $p_T>0.5$ GeV and/or small rapidities. At low values of $p_T$ and large rapidities the same deviation as above is observed. All protons including cluster fragment forming protons are treated as free protons which contributes to a more negative $v_1$.

\begin{figure}[t]	
\includegraphics[width=0.5\textwidth]{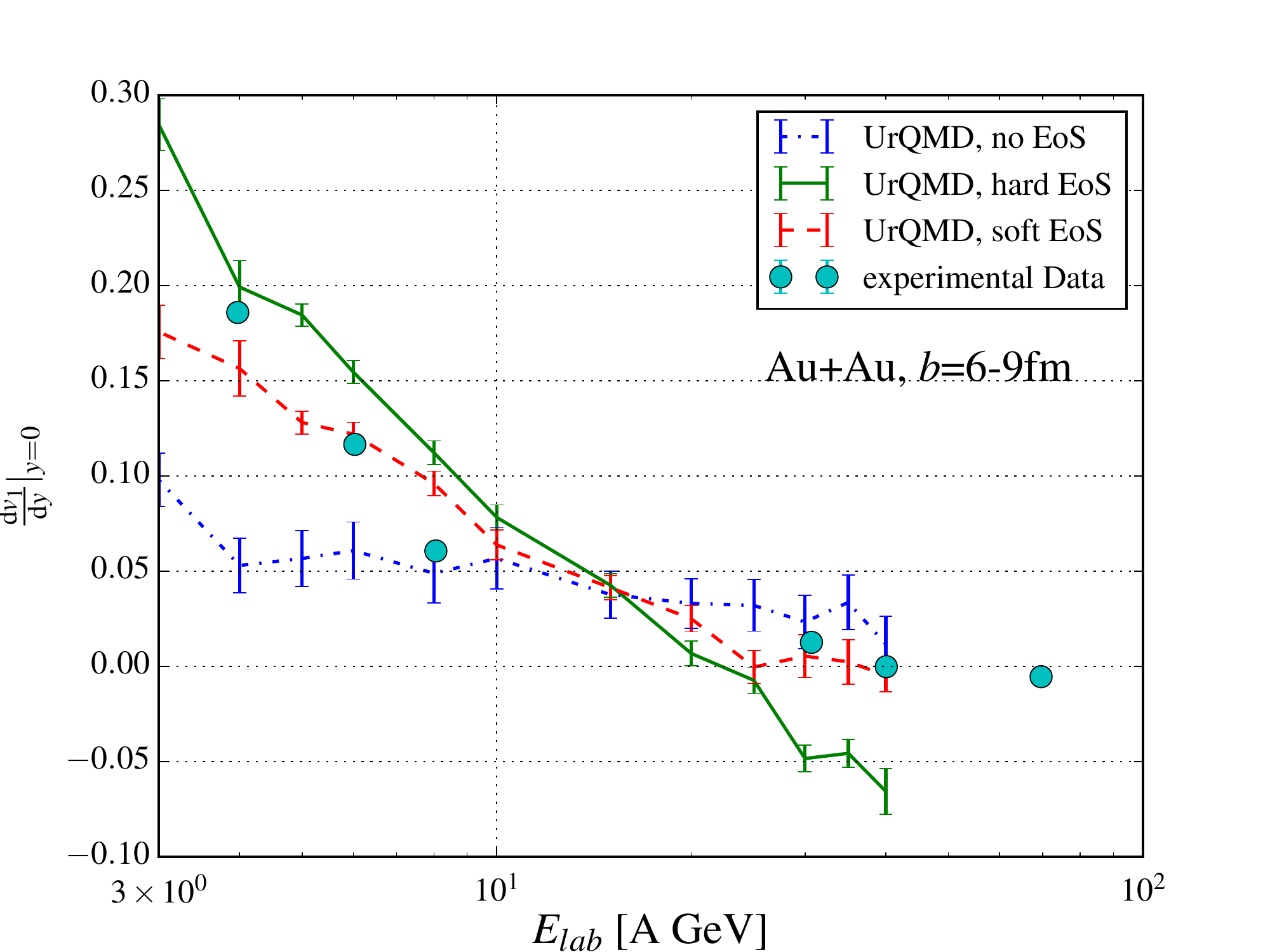}
\caption{[Color online] Zoom-in of the excitation function of the midrapidity slope parameters of $v_1$ of protons in Au+Au collisions (calculation: $b=6-9$ fm, data: 20\%-30\% centrality) for the cascade calculation and a soft and hard equation of state. The symbols denote the experimental data \cite{Alt:2003ab, Liu:2000am, Adamczyk:2014ipa}, the lines indicate the UrQMD calculations.}\label{f3b}
\end{figure}		
Let us next focus on the energy dependence of $v_1$. Due to momentum conservation the directed flow is exactly zero at $y=0$, and therefore one usually extracts the slope of $v_1$ with respect to the rapidity $y$ at $y=0$:

\begin{equation}\label{dv1dy}
\left. \frac{dv_1}{dy}\right|_{y=0}
\end{equation}

We calculate the slope from the $v_1$-bins  at $y=-0.1\pm 0.05$ and $y=0.1\pm 0.05$.

Figure \ref{f3} and \ref{f3b} show the excitation function of the midrapidity slope parameters of $v_1$ in Au+Au collisions (calculation: $b=6-9$ fm, data: 20\%-30\% centrality) for the cascade calculation, a soft and a hard equation of state. The symbols denote the experimental data, the lines indicate the UrQMD calculations.

A clear dependence of the directed flow slope on the equation of state is observed (which has already been found in previous publications \cite{Nara:2016hbg,Petersen:2006vm,Nara:2016phs,Ivanov:2014ioa}). The comparison with HADES data points in figure \ref{f2} suggested a rather hard equation of state.
In contrast, the data above $E_{\mathrm{lab}}>5$ A GeV tend to favor a soft EoS. In particular in the FAIR and RHIC BES region a softened EoS seems to describe the observed negative slope of $v_1$ better~\cite{Alt:2003ab,Adamczyk:2014ipa}.
In the absence of potentials $\left. \frac{dv_1}{dy}\right|_{y=0}$ remains positive, while in the cases with potentials it switches sign and becomes negative.
Such a behavior was predicted as a signal for a phase transition \cite{Csernai:1999nf,Brachmann:1999xt,Stoecker:2004qu}, which is not present in our current setup.

\subsection{Elliptic flow, $v_2$}

\begin{figure}[t]	
\includegraphics[width=0.5\textwidth]{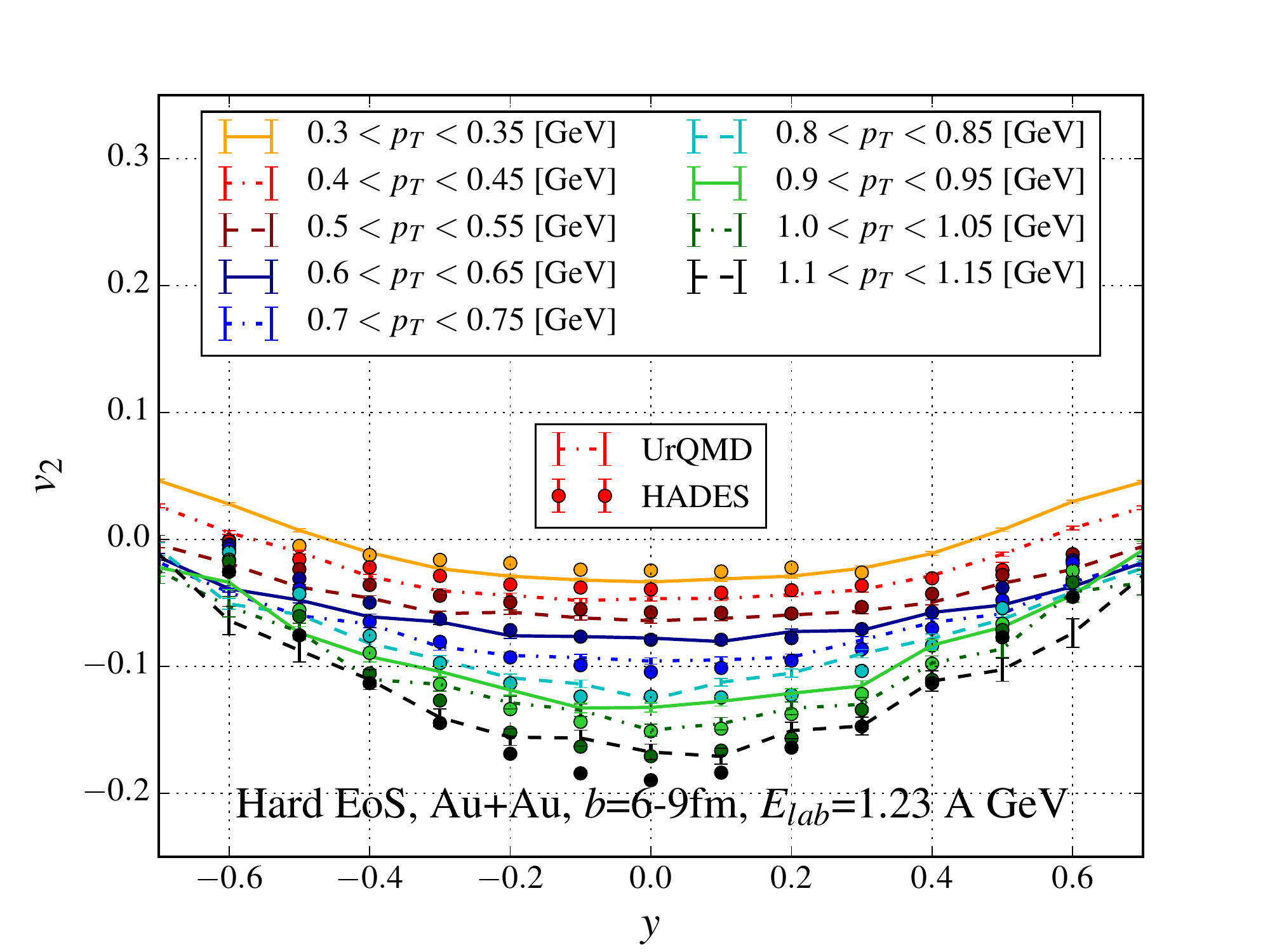}
\caption{[Color online] Elliptic flow of protons in Au+Au collisions (calculation: $b=6-9$ fm, data: 20\%-30\% centrality) as a function of rapidity and for various transverse momentum regions. The symbols denote the preliminary experimental data of the HADES collaboration~\cite{Kardan:2017knj}, the lines indicate the UrQMD calculations. 
}\label{f4}
\end{figure}		
\begin{figure}[t]	
\includegraphics[width=0.5\textwidth]{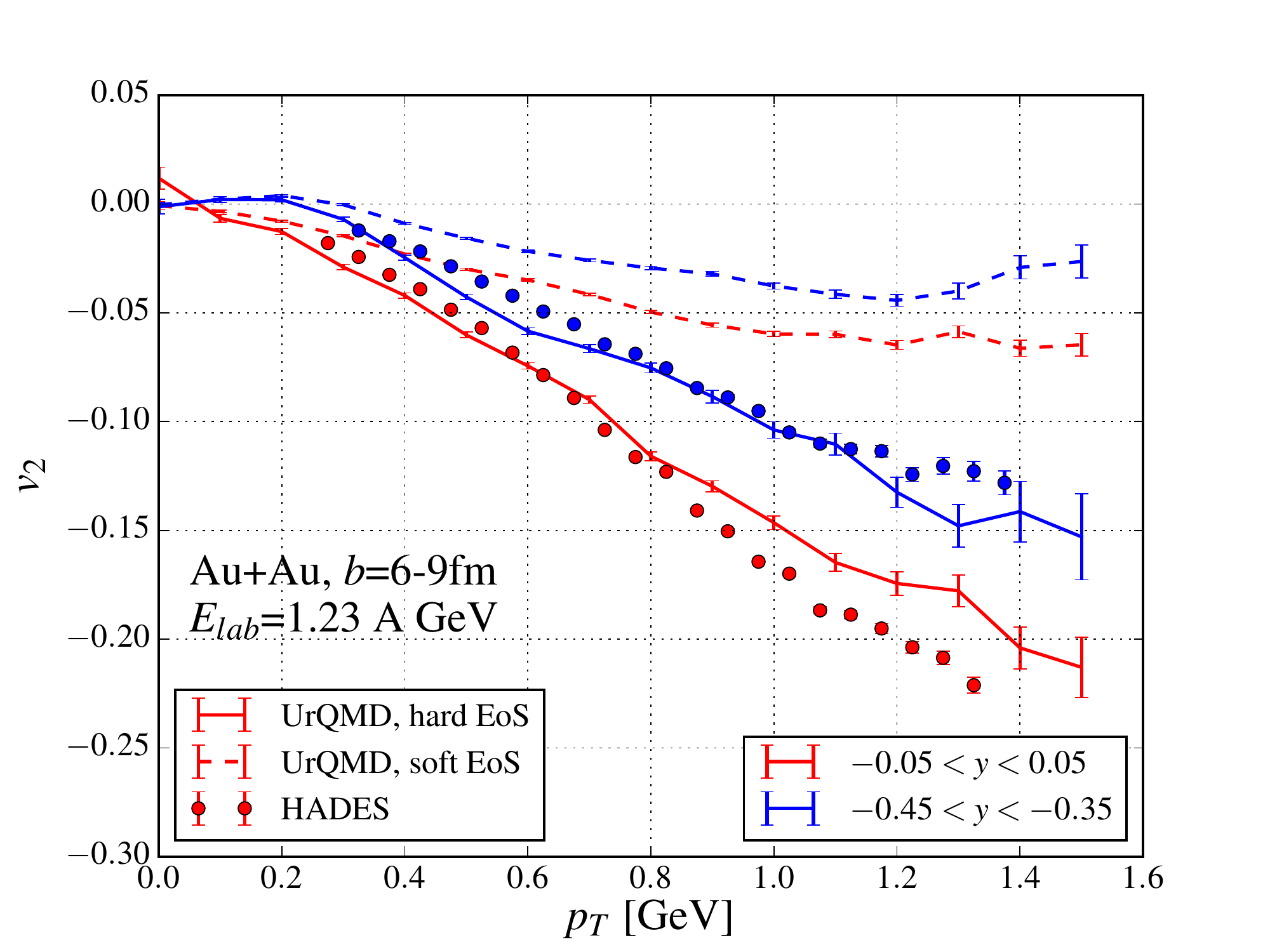}
\caption{[Color online] Elliptic flow of protons in Au+Au collisions (calculation: $b=6-9$ fm, data: 20\%-30\% centrality) as a function of transverse momentum for various rapidity regions. The symbols denote the preliminary experimental data of the HADES collaboration~\cite{Kardan:2017knj}, the lines indicate the UrQMD calculations. 
}\label{f5}
\end{figure}		

\begin{figure}[t]	
\includegraphics[width=0.5\textwidth]{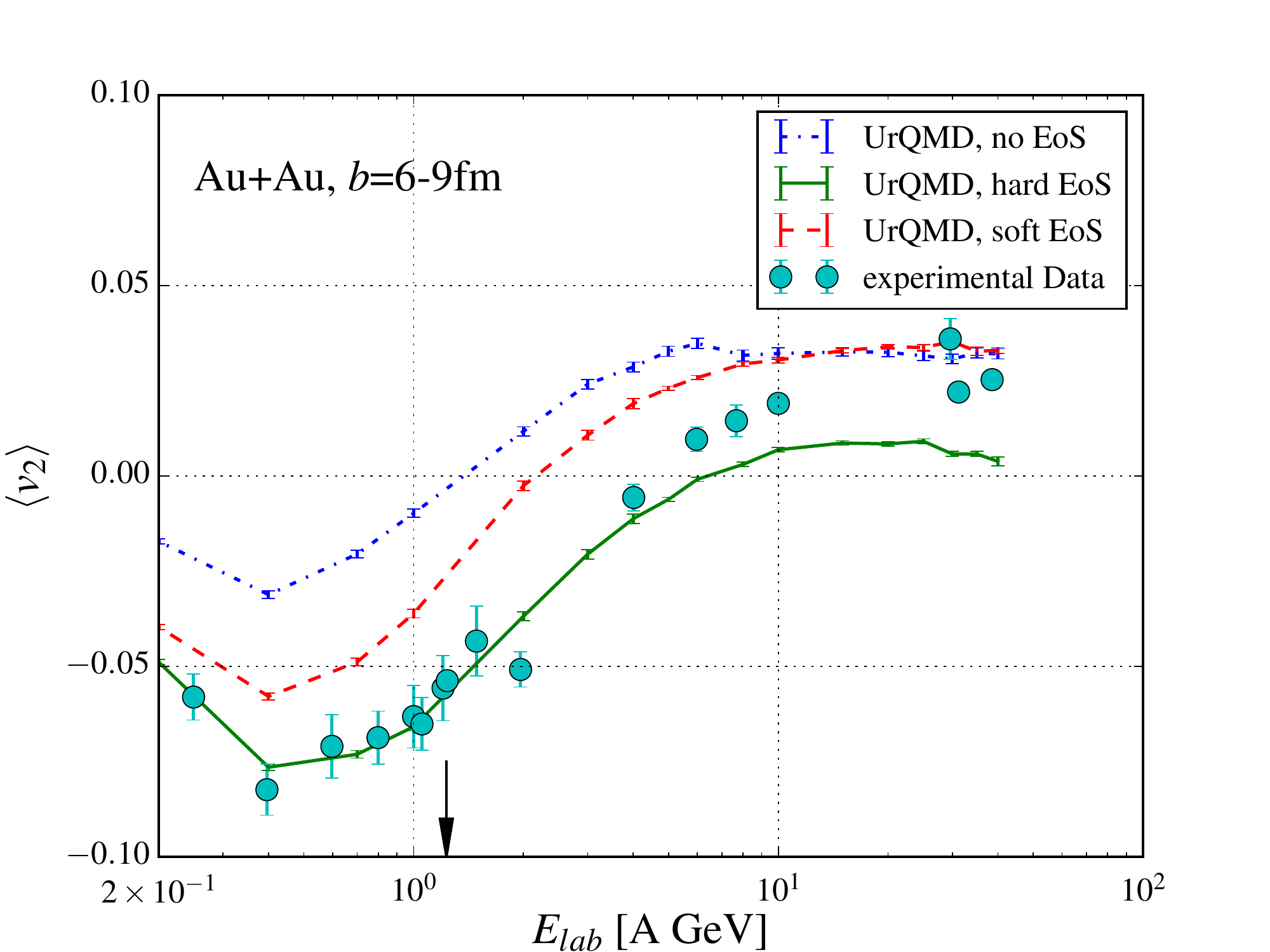}
\caption{[Color online] Excitation function of the elliptic flow of protons in Au+Au collisions (calculation: $b=6-9$ fm, data: 20\%-30\% centrality) for the cascade calculation and a soft and hard equation of state. The symbols denote the experimental data \cite{Andronic:2004cp,Adamczyk:2012ku,Pinkenburg:1999ya,BraunMunzinger:1998cg,Adamova:2002qx,Alt:2003ab}, the lines indicate the UrQMD calculations. The vertical line remarks the HADES energy.
}\label{f6}
\end{figure}		

The study of the elliptic flow has become a standard analysis tool to explore the properties of QCD-matter. At ultra-relativistic energies it allows to investigate e.g. the shear viscosity of the QGP stage. At lower energies as discussed here it sheds light on the equation of state of nuclear matter.
The elliptic flow at low beam energies is a result of the so called squeeze-out effect, where particles are blocked from being emitted in the reaction plane by the spectator nucleons and are emitted therefore mainly in the out-of-plane-direction. This leads to a negative value of the elliptic flow $v_2$ with respect to the reaction plane.
As the beam energy is increased, the spectators rapidly leave the collision zone and the emission of particles is now dominated by the initial pressure gradients of the ellipsoidal shape of the overlap region. In this case the elliptic flow coefficient turns positive, as particles in the reaction plane obtain a larger longitudinal flow velocity. 
Thus in the low energy regimes the elliptic flow is governed by an intricate interplay between the overlap stage and the subsequent expansion stage. 

Figure \ref{f4} shows the elliptic flow in Au+Au collisions (calculation: $b=6-9$ fm, data: 20\%-30\% centrality) as a function of rapidity and for various transverse momentum intervals at a fixed-target beam energy of 1.23 A GeV. The symbols denote the preliminary experimental data~\cite{Kardan:2017knj}, the lines indicate the UrQMD calculations with a hard EoS. 

For higher transverse momenta, the elliptic flow becomes more and more negative due to its growing $p_y$-component for higher $p_T$.

Figure \ref{f5} shows the elliptic flow in Au+Au collisions (calculation: $b=6-9$ fm, data: 20\%-30\% centrality) as a function of transverse momentum for various rapidity regions at a fixed-target beam-energy of 1.23 A GeV. The symbols denote the experimental data~\cite{Kardan:2017knj}, the lines indicate the UrQMD calculations. 

\begin{figure}[t]	
\includegraphics[width=0.5\textwidth]{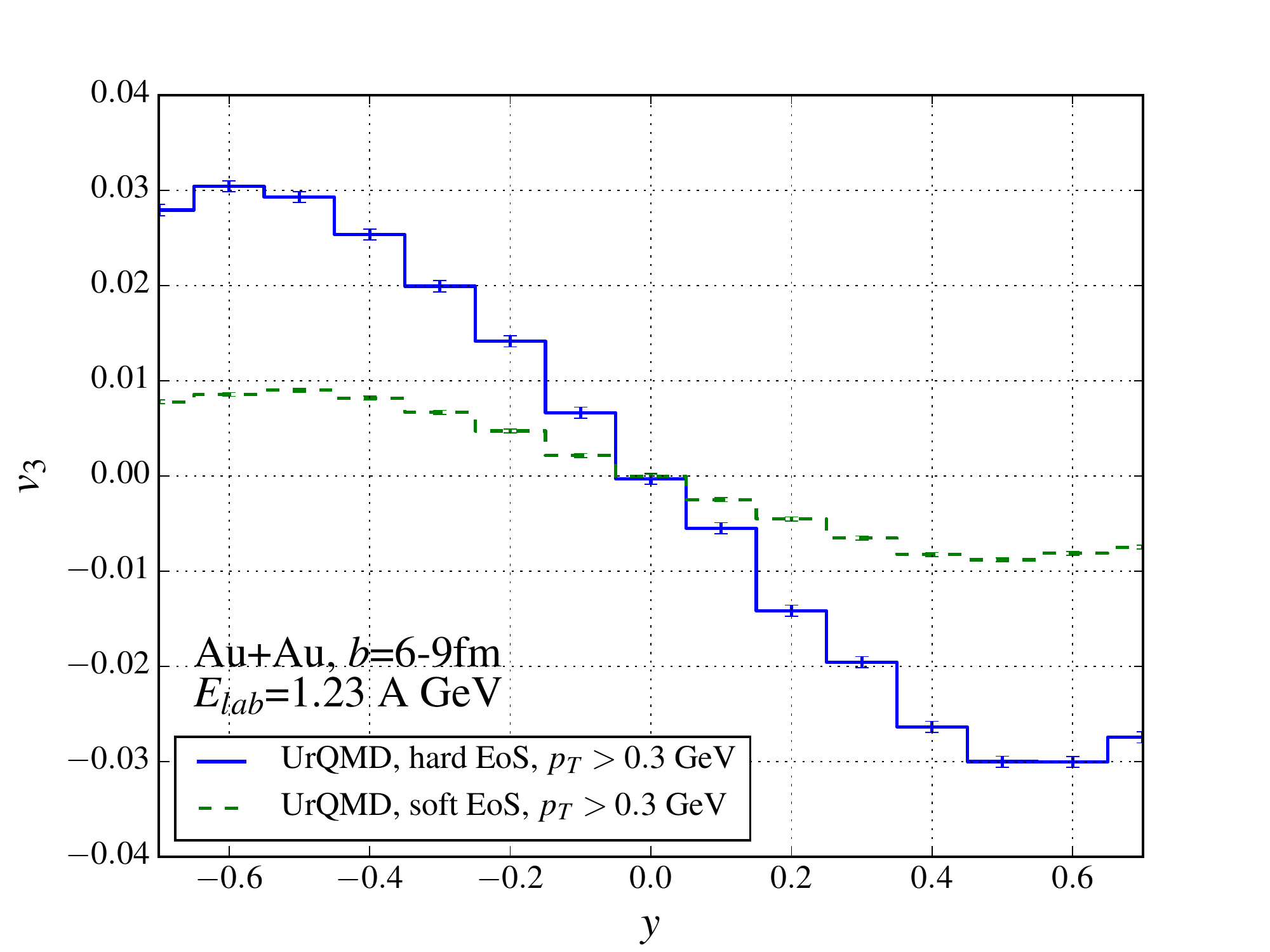}
\caption{[Color online] Triangular flow of protons in Au+Au collisions (calculation: $b=6-9$ fm, data: 20\%-30\% centrality) as a function of rapidity and for $p_T>0.3$ GeV. The lines indicate the UrQMD calculations for a hard and soft equation of state. 
}\label{f7}
\end{figure}		

For both rapidity windows one observes a strongly decreasing $v_2$ with higher $p_T$. Coming closer to midrapidity this behavior becomes even stronger. Again the calculations with a hard EoS result in a better description of preliminary HADES data.

Figure \ref{f6} shows the excitation function of the elliptic flow in mid-peripheral Au+Au collisions. Here we compare the cascade calculation and simulation with a soft and a hard equation of state. The symbols denote the experimental data, the lines indicate the UrQMD calculations. The black arrow denotes the HADES energy. The impact parameter varies between values of 6-9 fm. Note that the experimental data are not all for protons (sometimes charged particles are used) and use different centrality selections. Therefore a detailed comparison has to be done with caution.

Again, one observes a strong dependence of the flow on the equation of state, a similar behavior was also found in \cite{LeFevre:2016vpp} for IQMD calculations. For low energies till 0.4 A GeV one finds a decreasing flow which has its minimum at $E_{lab}=0.4$ A GeV for all scenarios, even the calculation in cascade mode. The decrease is due to the squeeze-out-effect~\cite{Hofmann:1976dy,Stoecker:1979mj,Stoecker:1980vf,Stoecker:1981pg}. For higher beam energies one observes an increase of the elliptic flow which slows down for intermediate energies, at which the sign of the elliptic flow has turned positive. In-plane emission becomes more and more preferable due to the higher energy and momenta of the nucleons. A similar behavior of the flow was also observed in~\cite{Petersen:2006vm} where additionally a momentum-dependent potential was included. It is important that, also for the elliptic flow, at higher beam energies $E_{\mathrm{lab}}> 5$ A GeV, a soft EoS is preferred by the data. The  softening can be observed in the same energy region like for the slope of the direct flow. This agreement is of great importance as it may indicate a formation of a mixed phase in this energy region.

\begin{figure}[t]	
\includegraphics[width=0.5\textwidth]{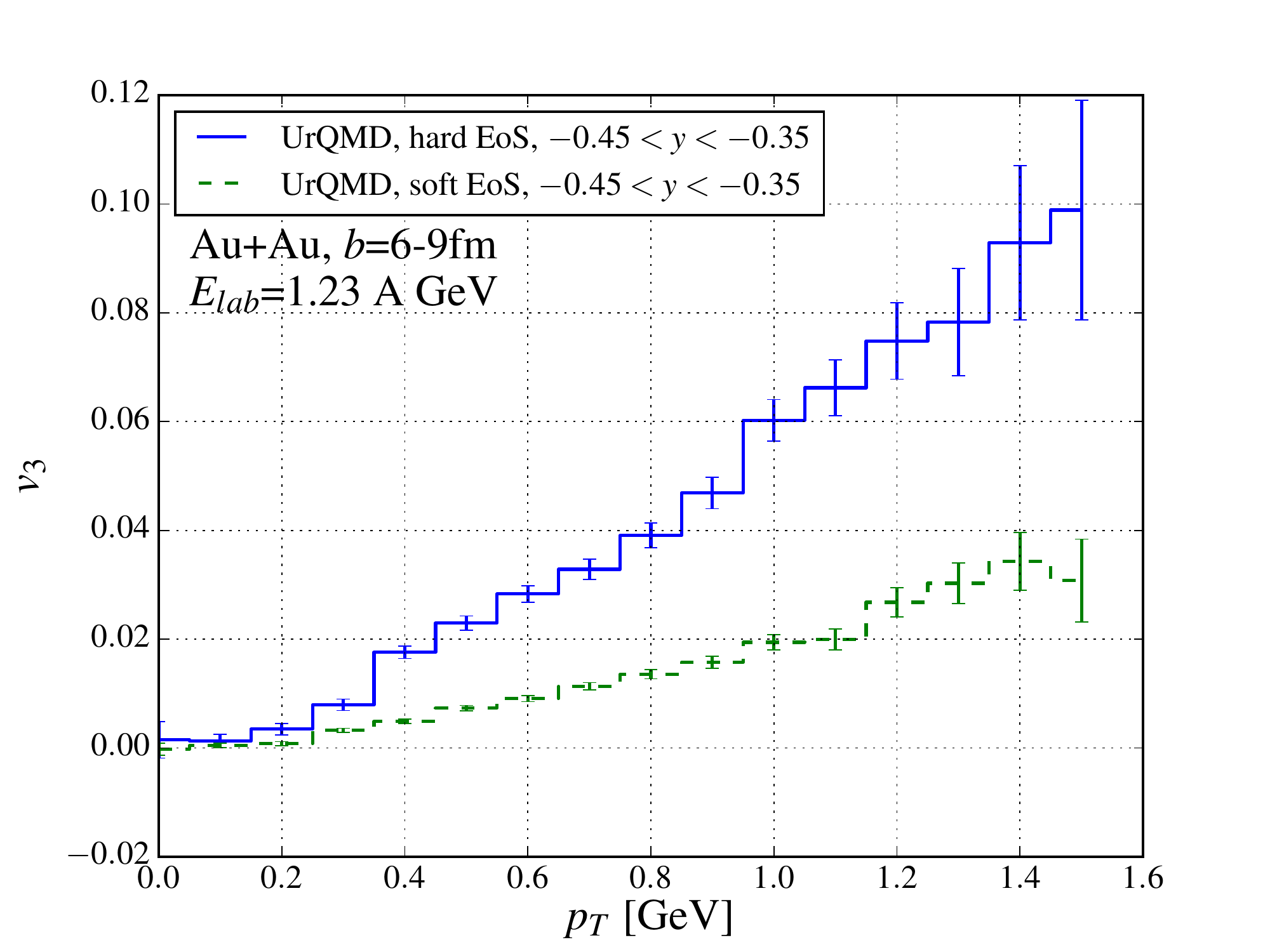}
\caption{[Color online] Triangular flow of protons in Au+Au collisions (calculation: $b=6-9$ fm, data: 20\%-30\% centrality) as a function of transverse momentum for $-0.45<y<-0.35$. The lines indicate the UrQMD calculations for a hard and soft equation of state. 
}\label{f8}
\end{figure}		

\subsection{Triangular flow, $v_3$}
The third Fourier coefficient, the triangular flow, has been mainly discussed as a consequence of initial state fluctuations (see e.g. \cite{Petersen:2010cw,Wang:2014boa,Teaney:2010vd}), which are uncorrelated with the reaction plane. This assumption was shown to be true for the highest beam energies available. As we will see now this interpretation is not any longer valid at low energies.
For the triangular flow studies, we evaluate $v_3$ with respect to the reaction plane. At very high energies it is clear that this procedure will result in $v_3=0$. At the low energies under investigation in the present paper, we may expect, however an interplay in the emission time structure together with the sizable $v_1$ component to yield a finite $v_3\neq 0$. 

Figure \ref{f7} shows the triangular flow in Au+Au collisions (calculation: $b=6-9$ fm, data: 20\%-30\% centrality) as a function of rapidity and for a transverse momentum $p_T>0.3$ GeV. The lines indicate the UrQMD calculations with a hard and soft EoS in comparison. From this result we can draw an important conclusion. We confirm a finite $v_3$ with respect to the reaction plane (at $y \ne 0 $) for low beam energies. Thus, at the energies discussed here, $v_3$ is correlated to the reaction plane.

\begin{figure}[t]	
\includegraphics[width=0.5\textwidth]{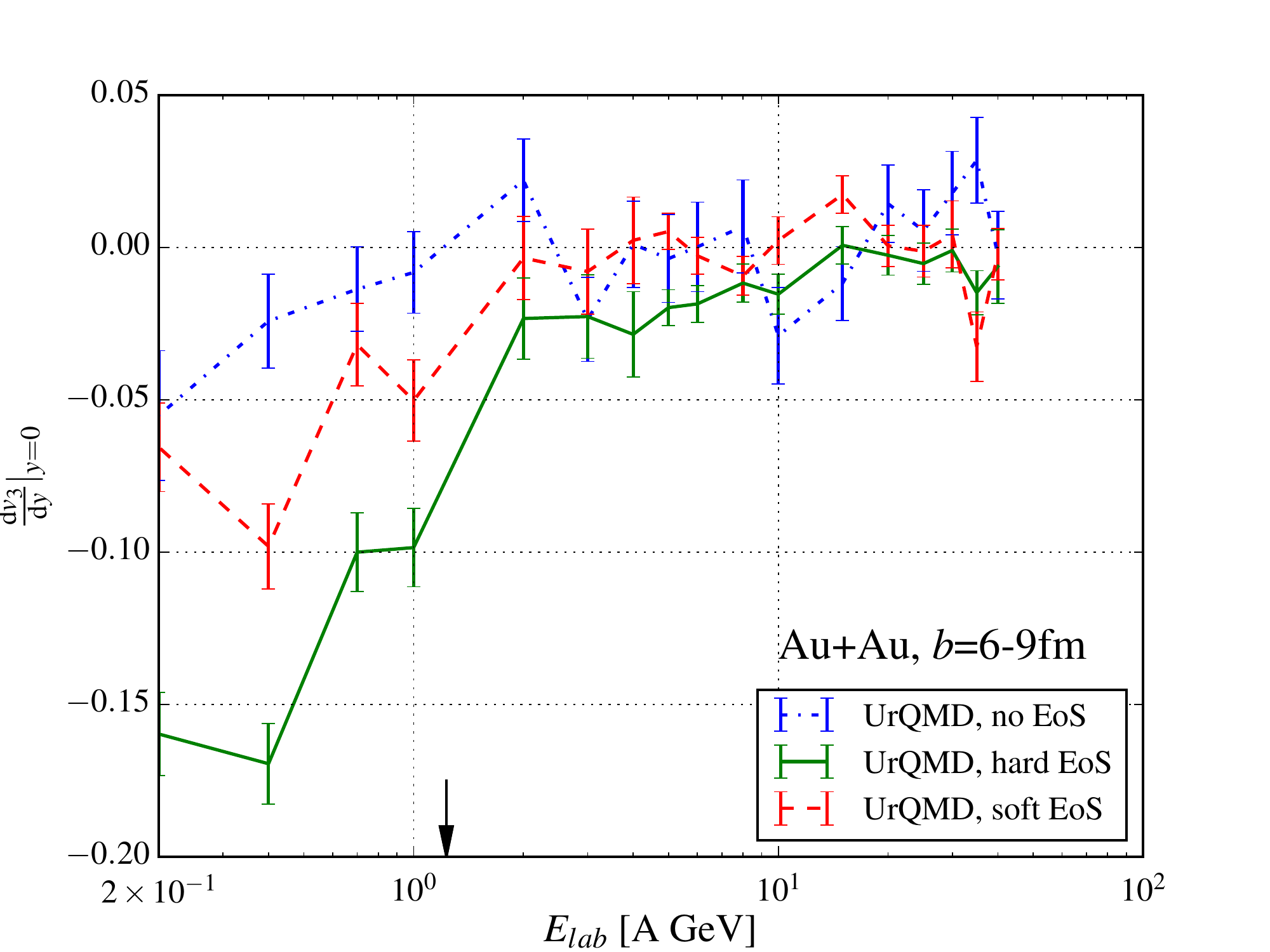}
\caption{[Color online] Excitation function of the triangular flow of protons in Au+Au collisions (calculation: $b=6-9$ fm, data: 20\%-30\% centrality) for the cascade calculation and a soft and hard equation of state. The arrow marks the HADES energy.
}\label{f9}
\end{figure}		

Figure \ref{f8} shows the triangular flow in Au+Au collisions (calculation: $b=6-9$ fm, data: 20\%-30\% centrality) as a function of transverse momentum for the backward rapidity region $-0.45<y<-0.35$. Here, the soft and the hard EoS are compared. A clear dependence of the EoS is observed.
Since $v_3$ appears due to a combination of the geometry and space-time-dynamics, we expect it to be even more sensitive to the strength of the potential than $v_1$ and $v_2$. $v_3$ provides a crucial test for the modeling of the dynamics of the system.

As for the $v_1$, the triangular flow vanishes at midrapidity due to momentum conservation, so we extract the slope with respect to the rapidity:\\
\begin{equation}
\left. \frac{dv_3}{dy}\right|_{y=0}
\end{equation}

Again, we calculate the slope of $v_3$ from the bins at $y=-0.1\pm 0.05$ and $y=0.1\pm 0.05$.

Figure \ref{f9} shows the excitation function of the slope of the triangular flow in Au+Au collisions ($b$=6-9 fm) for the cascade calculation and a soft and hard equation of state. 
For all cases a non-zero $v_3$ with respect to the reaction plane is observed up to beam energies of $E_{\mathrm{lab}}<5$ A GeV. Its magnitude depends strongly on the EoS. This first measurement of $v_3\neq 0$ with respect to the reaction-plane indicates an interplay of initial stage and expansion stage of the system. Thus, it is not only a consequence of initial state fluctuations. At low energies this makes a separation of both stages impossible like many models, e.g. hydrodynamics, assume.

\section{Summary}
We presented a transport model study of the preliminary HADES data for Au+Au reaction at a beam energy of 1.23 A GeV. The UrQMD model provides a very good description of the available data, if a hard equation of state is employed. 

Investigating the beam energy dependence of $v_1$ and $v_2$ we found that the data consistently favors a hard EoS for $E_{\mathrm{lab}}<5$ A GeV and a softening of the EoS for higher beam energies, consistent with findings in \cite{Nara:2016hbg,Nara:2017qcg}. This behavior can be studied in more detail at the upcoming FAIR facility.

For the first time we predicted a finite triangular flow, $v_3$, with respect to the reaction plane in the SIS18-SIS100 energy range. This observation is in striking contrast to all results obtained at higher energies, where only the $v_3$ with respect to the 3rd order event plane is finite, which sensitive to initial state fluctuations, not correlated to the reaction plane. The triangular flow is sensitive to the equation of state and can serve as a new tool to explore the time dependence of the pressure during the collision. The finite $v_3$ indicates that the different stages of the reaction can not be separated into an initial stage and an expansion stage, which is assumed in many hybrid approaches and hydrodyanmics models. Therefore, it questions the applicability of many currently used models for the exploration of the phase transition at FAIR energies. A possible solution of this problem could be full 3+1 dimensional multi-fluid approaches that can cover the whole evolution of the system from the start to the end, including a phase transition.


\section{Acknowledgments}
The authors thank Behruz Kardan, Manuel Lorenz and Christoph Blume for helpful and inspiring discussions as well as Yongjia Wang for his help with the nuclear potential parameters.
The computational resources were provided by the LOEWE Frankfurt Center for Scientific Computing (LOEWE-CSC). This work was supported by  HIC for FAIR and in the framework of COST Action CA15213 THOR.   


\begin{thebibliography}{100}





\bibitem{Kardan:2016uog} 
  B.~Kardan,
  J.\ Phys.\ Conf.\ Ser.\  {\bf 742}, no. 1, 012008 (2016).

\bibitem{McLerran:2008ux}
  L.~McLerran,
  arXiv:0808.1057 [hep-ph].

\bibitem{Ruester:2006aj}
  S.~B.~Ruester, V.~Werth, M.~Buballa, I.~A.~Shovkovy and D.~H.~Rischke,
  arXiv:nucl-th/0602018.

\bibitem{Poskanzer:1998yz}
  A.~M.~Poskanzer and S.~A.~Voloshin,
  Phys.\ Rev.\  C {\bf 58}, 1671 (1998).

\bibitem{Stoecker:1979mj}
  H.~Stoecker, J.~A.~Maruhn and W.~Greiner,
  Z.\ Phys.\  A {\bf 290}, 297 (1979).

\bibitem{Hofmann:1976dy}
  J.~Hofmann, H.~Stoecker, U.~W.~Heinz, W.~Scheid and W.~Greiner,
  Phys.\ Rev.\ Lett.\  {\bf 36}, 88 (1976).

\bibitem{Stoecker:1986ci}
  H.~Stoecker and W.~Greiner,
  Phys.\ Rept.\  {\bf 137}, 277 (1986).

\bibitem{Nara:2016hbg} 
  Y.~Nara, H.~Niemi, J.~Steinheimer and H.~Stoecker,
  Phys.\ Lett.\ B {\bf 769}, 543 (2017)

\bibitem{Nara:2016phs} 
  Y.~Nara, H.~Niemi, A.~Ohnishi and H.~Stoecker,
  Phys.\ Rev.\ C {\bf 94}, no. 3, 034906 (2016).
  
\bibitem{Ivanov:2014ioa} 
  Y.~B.~Ivanov and A.~A.~Soldatov,
  Phys.\ Rev.\ C {\bf 91}, no. 2, 024915 (2015)
  [arXiv:1412.1669 [nucl-th]].
  
\bibitem{Petersen:2006vm}
  H.~Petersen, Q.~Li, X.~Zhu and M.~Bleicher,
  Phys.\ Rev.\  C {\bf 74}, 064908 (2006).
  
\bibitem{Nara:2017qcg} 
  Y.~Nara, H.~Niemi, A.~Ohnishi, J.~Steinheimer, X.~Luo and H.~Stoecker,
  arXiv:1708.05617 [nucl-th].

\bibitem{Isse:2005nk}
  M.~Isse, A.~Ohnishi, N.~Otuka, P.~K.~Sahu and Y.~Nara,
  Phys.\ Rev.\  C {\bf 72}, 064908 (2005).

\bibitem{Voloshin:2002wa}
  S.~A.~Voloshin,
  Nucl.\ Phys.\  A {\bf 715}, 379 (2003).

\bibitem{Snellings:2011sz}
  R.~Snellings,
  New J.\ Phys.\  {\bf 13}, 055008 (2011).

\bibitem{Ollitrault:1997vz}
  J.~Y.~Ollitrault,
  Nucl.\ Phys.\  A {\bf 638}, 195 (1998).

\bibitem{Yan:2013laa} 
  L.~Yan and J.~Y.~Ollitrault,
  Phys.\ Rev.\ Lett.\  {\bf 112}, 082301 (2014).

\bibitem{Retinskaya:2012ky}
  E.~Retinskaya, M.~Luzum and J.~Y.~Ollitrault,
  Phys.\ Rev.\ Lett.\  {\bf 108}, 252302 (2012).

\bibitem{Yin:2017qhg} 
  X.~Yin, C.~M.~Ko, Y.~Sun and L.~Zhu,
  Phys.\ Rev.\ C {\bf 95}, no. 5, 054913 (2017).

\bibitem{Adam:2016nfo} 
  J.~Adam {\it et al.} [ALICE Collaboration],
  JHEP {\bf 1609}, 164 (2016).

\bibitem{Esumi:2017qof} 
  S.~Esumi,
  EPJ Web Conf.\  {\bf 141}, 05001 (2017).

\bibitem{He:2017qsk} 
  L.~He [STAR Collaboration],
  Nucl.\ Phys.\ A {\bf 967}, 616 (2017).

\bibitem{Krzewicki:2011ee}
  M.~Krzewicki  [ALICE Collaboration],
  J.\ Phys.\  {\bf G38}, 124047 (2011).

\bibitem{Alver:2010gr}
  B.~Alver and G.~Roland,
  Phys.\ Rev.\  C {\bf 81}, 054905 (2010)
  [Erratum-ibid.\  C {\bf 82}, 039903 (2010)].

\bibitem{Bass:1998ca}
  S.~A.~Bass {\it et al.},
  Prog.\ Part.\ Nucl.\ Phys.\  {\bf 41}, 255 (1998)
  [Prog.\ Part.\ Nucl.\ Phys.\  {\bf 41}, 225 (1998)].

\bibitem{Bleicher:1999xi}
  M.~Bleicher {\it et al.},
  J.\ Phys.\  {\bf G25}, 1859 (1999).

\bibitem{Graef:2014mra} 
  G.~Graef, J.~Steinheimer, F.~Li and M.~Bleicher,
  Phys.\ Rev.\ C {\bf 90}, 064909 (2014).

\bibitem{Patrignani:2016xqp} 
  C.~Patrignani {\it et al.} [Particle Data Group],
  Chin.\ Phys.\ C {\bf 40}, no. 10, 100001 (2016).

\bibitem{Bleicher:2005ti}
  M.~Bleicher, E.~Bratkovskaya, S.~Vogel and X.~Zhu,
  J.\ Phys.\  {\bf G31}, S709 (2005).

\bibitem{Hartnack:1997ez}
  C.~Hartnack, R.~K.~Puri, J.~Aichelin, J.~Konopka, S.~A.~Bass, H.~Stoecker and W.~Greiner,
  Eur.\ Phys.\ J.\  A {\bf 1}, 151 (1998).

\bibitem{Li:2005gfa}
  Q.~f.~Li, Z.~x.~Li, S.~Soff, M.~Bleicher and H.~Stoecker,
  J.\ Phys.\  {\bf G32}, 151 (2006).

\bibitem{Li:2007yd}
  Q.~Li, M.~Bleicher and H.~Stoecker,
  Phys.\ Lett.\  B {\bf 659}, 525 (2008).
  
\bibitem{Li:2005zza} 
  Q.~f.~Li, Z.~x.~Li, S.~Soff, R.~K.~Gupta, M.~Bleicher and H.~Stoecker,
  J.\ Phys.\ G {\bf 31}, 1359 (2005)
  [nucl-th/0507068].

\bibitem{Kardan:2017knj} 
  B.~Kardan [HADES Collaboration],
  Nucl.\ Phys.\ A {\bf 967}, 812 (2017).

\bibitem{LeFevre:2016vpp} 
  A.~Le Fevre, Y.~Leifels, C.~Hartnack and J.~Aichelin,
  arXiv:1611.07500 [nucl-th].

\bibitem{Alt:2003ab}
  C.~Alt {\it et al.}  [NA49 Collaboration],
  Phys.\ Rev.\  C {\bf 68}, 034903 (2003).
  
\bibitem{Liu:2000am}
  H.~Liu {\it et al.}  [E895 Collaboration],
  Phys.\ Rev.\ Lett.\  {\bf 84}, 5488 (2000).

\bibitem{Adamczyk:2014ipa} 
  L.~Adamczyk {\it et al.} [STAR Collaboration],
  Phys.\ Rev.\ Lett.\  {\bf 112}, no. 16, 162301 (2014).

\bibitem{Csernai:1999nf} 
  L.~P.~Csernai and D.~Rohrich,
  Phys.\ Lett.\ B {\bf 458}, 454 (1999)

\bibitem{Brachmann:1999xt} 
  J.~Brachmann, S.~Soff, A.~Dumitru, H.~Stoecker, J.~A.~Maruhn, W.~Greiner, L.~V.~Bravina and D.~H.~Rischke,
  Phys.\ Rev.\ C {\bf 61}, 024909 (2000)

\bibitem{Stoecker:2004qu} 
  H.~Stoecker,
  Nucl.\ Phys.\ A {\bf 750}, 121 (2005)

\bibitem{Andronic:2004cp}
  A.~Andronic {\it et al.}  [FOPI Collaboration],
  Phys.\ Lett.\  B {\bf 612}, 173 (2005).

\bibitem{Adamczyk:2012ku}
  L.~Adamczyk {\it et al.}  [STAR collaboration],
  Phys.\ Rev.\  C {\bf 86}, 054908 (2012).

\bibitem{Pinkenburg:1999ya}
  C.~Pinkenburg {\it et al.}  [E895 Collaboration],
  Phys.\ Rev.\ Lett.\  {\bf 83}, 1295 (1999).

\bibitem{BraunMunzinger:1998cg}
  P.~Braun-Munzinger and J.~Stachel,
  Nucl.\ Phys.\  A {\bf 638}, 3 (1998).

\bibitem{Adamova:2002qx}
  D.~Adamova {\it et al.}  [CERES Collaboration],
  Nucl.\ Phys.\  A {\bf 698}, 253 (2002).


\bibitem{Stoecker:1980vf}
  H.~Stoecker, J.~A.~Maruhn and W.~Greiner,
  Phys.\ Rev.\ Lett.\  {\bf 44}, 725 (1980).

\bibitem{Stoecker:1981pg}
  H.~Stoecker {\it et al.},
  Phys.\ Rev.\  C {\bf 25}, 1873 (1982).

\bibitem{Petersen:2010cw} 
  H.~Petersen, G.~Y.~Qin, S.~A.~Bass and B.~Muller,
  Phys.\ Rev.\ C {\bf 82}, 041901 (2010).
  
\bibitem{Wang:2014boa} 
  J.~Wang, Y.~G.~Ma, G.~Q.~Zhang and W.~Q.~Shen,
  Phys.\ Rev.\ C {\bf 90}, no. 5, 054601 (2014)

\bibitem{Teaney:2010vd} 
  D.~Teaney and L.~Yan,
  Phys.\ Rev.\ C {\bf 83}, 064904 (2011).
 
\end{thebibliography}
\end{document}